\documentclass[
  aps,
  prapplied,
  reprint,
  amsmath,
  amssymb,
  nofootinbib,
  groupedaddress,
  floatfix,
  longbibliography
]{revtex4-2}

\usepackage[utf8]{inputenc}
\usepackage{graphicx}
\usepackage{braket}
\usepackage{multirow}
\usepackage{booktabs}
\usepackage{hyperref}
\usepackage{balance}
\usepackage{xcolor}
\usepackage{float}
\usepackage{subcaption}
\usepackage{amsthm}
\usepackage{dsfont}
\usepackage{tcolorbox}
\usepackage{tabularx}
\usepackage{enumitem} 
\usepackage{tikz}
\usetikzlibrary{arrows.meta, positioning, shapes.geometric}

\usepackage{tcolorbox}
\tcbuselibrary{skins, breakable}
\usepackage{lipsum}
\usepackage{ulem}
\usepackage{color}
\usepackage[dvipsnames]{xcolor}

\usepackage{algorithm} 
\usepackage{algpseudocode}

\tcbuselibrary{skins, breakable}

\begin{document}
\title{Benchmarking Quantum Extreme Learning based on Gaussian Boson Sampling}

\author{Daniel Montesinos, Gian Luca Giorgi, Roberta Zambrini}
\affiliation{IFISC, Institute for Cross-Disciplinary Physics and Complex Systems (UIB-CSIC)\\
Campus Universitat Illes Balears, 07122, Palma de Mallorca, Spain}

\begin{abstract}

Reservoir models offer a hardware-efficient learning paradigm for noisy intermediate-scale quantum devices by exploiting untrained quantum dynamics as a fixed feature map and restricting optimization to a simple classical readout layer. We propose a quantum extreme learning machine implemented using gaussian boson sampling and an encoding strategy that achieves high classification accuracy while reducing optical resource requirements. Classical inputs are jointly encoded in the squeezing parameters and in the interferometer unitary, enabling sampling-based, highly nonlinear feature maps while leveraging large-scale GBS output statistics, which are conjectured to be classically intractable. We systematically compare multiple families of quantum features accessible in the same setup and find that photon-number sampling probabilities provide the best performance, consistent with their higher effective feature dimensionality. Finally, we benchmark against classical nonlinear baselines and analyse robustness under noisy scenarios, showing competitive performance with fewer trainable parameters and indicating practical promise for near-term photonic implementations.

\end{abstract}
\maketitle
\section{Introduction}

Machine learning is central to modern technology, with supervised classification powering applications in various fields \cite{lecunDeepLearning2015,minaeeDeepLearningbasedText2022}. However, it usually relies on models with high computational and energy costs due to large parameter counts and iterative training \cite{schwartzGreenAI2020}. This motivates reservoir computing frameworks, including Extreme Learning Machines (ELM's), where a fixed nonlinear transformation generates features and training is reduced to a lightweight linear readout \cite{huangExtremeLearningMachine2006,lukoseviciusReservoirComputingApproaches2009}. A central challenge in this framework is the construction of feature maps that are both highly expressive and resource-efficient. In this context, quantum systems offer a promising alternative, as their intrinsic dynamics can naturally generate complex, high-dimensional transformations of input data \cite{biamonteQuantumMachineLearning2017, fujiiHarnessingDisorderedEnsembleQuantum2017,mujalOpportunitiesQuantumReservoir2021}. Recent studies have established key theoretical aspects of Quantum Extreme Learning Machines (QELMs) \cite{innocentiPotentialLimitationsQuantum2023b, xiongFundamentalAspectsQuantum2025}, with recent applications on a wide range of quantum platforms and learning tasks \cite{sakuraiQuantumExtremeReservoir2022, delorenzisHarnessingQuantumExtreme2025, vetranoStateEstimationQuantum2025a, lauModularQuantumExtreme2026}. 

Quantum photonics offers a natural route to such architectures \cite{Pelucchi2022IntegratedPhotonicQuantumTech,asavanantGenerationTimedomainmultiplexedTwodimensional2019,Shastri2021PhotonicsAI,labay-moraNeuralNetworksQuantum2024,brunnerRoadmapNeuromorphicPhotonics2025,aghaeeradScalingNetworkingModular2025a, Ulrik2025, Paparelle2026}, and in this context, Gaussian Boson Sampling (GBS) is a particularly attractive platform \cite{2017GBS}. In GBS, squeezed states propagate through a linear optical interferometer and photon number measurements are performed at the output. The evolution remains Gaussian, while the computationally rich structure appears in the measurement statistics, with output probabilities governed by Hafnians \cite{2017GBS,detailedGBS}. Under well-motivated complexity theoretic assumptions, GBS is widely believed to be classically hard to simulate even approximately \cite{2017GBS,aaronsonComputationalComplexityLinear2011}, and large-scale experiments have reported results consistent with operating in regimes challenging for classical computation \cite{spagnoloExperimentalValidationPhotonic2014, Sciarrino2015, wangBosonSampling202019,zhongQuantumComputationalAdvantage2020,zhongPhaseProgrammableGaussianBoson2021,madsenQuantumComputationalAdvantage2022a,dengGaussianBosonSampling2023, eickmannBridgingChemistryGaussian2026}. 

In particular, GBS platforms now span an impressive range of physical scales and encoding strategies. A large-scale device with 8176 spatio-temporal optical modes has been demonstrated \cite{liu2025}, and used for image classification via sampling probabilities \cite{gong2025}. In a complementary direction, an experiment with more than 400 optical modes in the frequency domain has also been employed for classification, but using the covariance matrix of photon correlations as the feature representation \cite{cimini2025}. These demonstrations highlight that classification with GBS can be approached either through sample-based features from photon-pattern outcomes or through average photon-number correlations. Importantly, promising results are not limited to very large photonic systems. Smaller GBS setups with 16 spatial optical modes have been used experimentally to build feature extraction models and quantum generative adversarial networks \cite{zhu2025}, indicating that useful learning primitives can emerge even with a modest number of modes. In that setting, it was sufficient to measure photon coincidence events corresponding to 2-photon and 4-photon outcomes, reinforcing the idea that carefully chosen measurement statistics can provide expressive features without requiring extreme hardware scaling.

Closely related photonic  efforts in classification tasks have also been pursued outside the GBS input model by using Fock state light instead of squeezed states. A scheme using modes that combine spatial and polarisation degrees of freedom has been proposed for image classification \cite{Nerenberg:25}, while approaches relying only on spatial modes have been investigated both in numerical simulations \cite{Sakurai:25} and experimentally \cite{rambach2025}. Other boson sampling-based approaches have also been used to build quantum support vector machines in a small photonic integrated processor \cite{Yin2025} and to use adaptive circuits as the core for computing kernel distances \cite{Hoch2025}. Furthermore, beyond standard boson-sampling architectures, photonic platforms have also been explored for ELM, including approaches based on the orbital angular momentum and polarization of single photons \cite{Suprano2024}. Additionally, a Quantum Reservoir Computing (QRC) protocol has recently been experimentally demonstrated in continuous-variable photonics for forecasting \cite{Paparelle2026} and also in the discrete case  \cite{Mirela2025}.

Despite recent progress, it remains unclear how to design resource-efficient photonic learning models that maximize feature expressivity without increasing hardware complexity. In this work, we make four main contributions. (i) We introduce a GBS classification model with a new encoding strategy that jointly embeds data into the input squeezing parameters and the interferometer unitary, and we show the performance improvement offered by this design. (ii) We benchmark, side by side, multiple quantum feature representations obtainable from different measurements using the same photonic setup and quantify the trade-off between accuracy and feature dimension. (iii) Our simulations suggest that competitive classification performance can be achieved with as few as 12 spatial modes, paving the way to a potentially more resource-efficient operation. Finally,
(iv) we assess robustness under noisy settings and benchmark against strong classical baselines, including nonlinear methods, for a fair comparison.

\section{QELM model}
In standard GBS, an $M$-mode interferometer is fed with single-mode squeezed vacuum states, $\hat S(r_i,\phi_i)\ket{0}$ with $S(r_i,\phi_i)=\mathrm{exp}(r_i(e^{-i\phi_i} \hat{a_i}^2 - e^{-i\phi_i} \ \hat{a_i}^{\dagger2})/2)$ where $r_i\ge 0$ and $\phi_i\in[0,2\pi)$ denote the squeezing strength and phase of each squeezed mode $i = 1, 2,\dots, M$. The optical field then propagates through a linear optical network described by a complex unitary $U\in\mathbb C^{M\times M}$, implemented with beam splitters and phase shifters, and often chosen at random. Photon Number Resolving (PNR) detection at the output produces samples with total photon number $N_p$, which is restricted to even values since inputs are squeezed vacuum (photons generated in pairs) and the interferometer is passive. The probability of each output Fock pattern depends jointly on the interferometer matrix elements and the squeezing parameters through the Hafnian, yielding a highly nonlinear dependence that is widely believed to be classically hard to compute in general. A detailed description of the GBS formalism is given in \cite{detailedGBS}, while \cite{Oh:25} offers a review of both its theoretical foundations and experimental implementations.

\begin{figure*}[t]
    \centering
    \includegraphics[width=\textwidth]{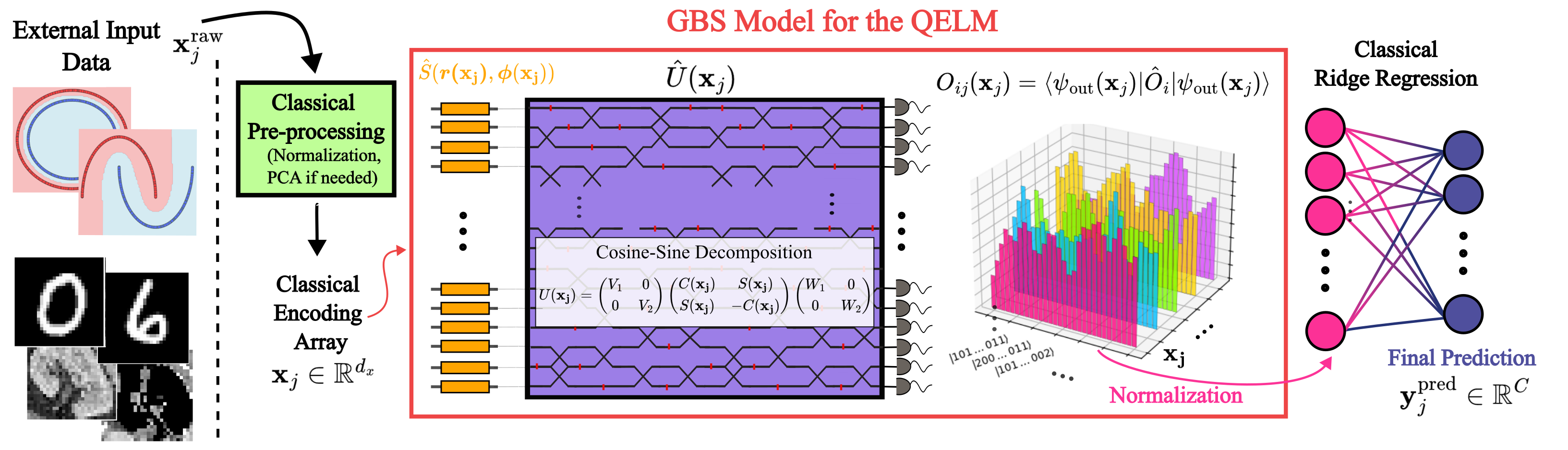}
    \caption{Schematic overview of the QELM based on GBS. External input samples $\mathbf{x}_j^{\text{raw}}$ undergo classical preprocessing (normalization and, when required, reduction via PCA) to generate the classical feature vector $\mathbf{x}_j \in \mathbb{R}^{d_x}$. These features modulate the squeezing strengths and phases applied to each of the $M$ optical modes, producing an input-dependent Gaussian state. This state is then processed by a linear interferometer $\hat{U}(\mathbf{x}_j)$, and photon-number-resolving detection is used to extract a set of quantum features $O_{ij}(\mathbf{x_j})$, corresponding to the probabilities of selected Fock outcomes. Finally, the resulting feature vector is fed into a classical Ridge regression layer, which produces the prediction $\mathbf{y}^{\mathrm{pred}}_j \in \mathbb{R}^{C}$ where $C$ is the number of distinct classes in the task.
}\label{fig:scheme}
\end{figure*}

The quantum model proposed in this work is a Quantum Extreme Learning Machine (QELM) based on the GBS setup. By encoding the input in both the squeezing parameters and the interferometer matrix, the input is mapped nonlinearly in the GBS output statistic. The goal is to perform multiclass classification efficiently using QELM. Given a classification labelled dataset $\{(\mathbf{x}^{\mathrm{raw}}_j,\mathbf{y}_j)\}$, with one-hot\footnote{One-hot encoding maps class index ${k_j}$ to $\mathbf{y}_j = \mathbf{e}_{k_j} \in \mathbb{R}^C$, where $\{\mathbf{e}_{k_j}\}$ are the canonical basis vectors and $C$ is the number of classes.} encoded targets $\mathbf{y}_j$, we process each input $\mathbf{x}^{\mathrm{raw}}_j$ through the optical circuit to predict its label. Three crucial aspects of the QELM design are the input encoding, the choice of output observables used as quantum features, and the training procedure. Figure~\ref{fig:scheme} summarizes the workflow, and the remainder of this section details each of these aspects.

\subsection{Input Encoding} \label{sec:Input Encoding}

The choice of input encoding is central in any QELM as well as QRC, as it determines how classical data is transformed into a nonlinear quantum feature space \cite{nokkala2021, schuldEffectDataEncoding2021, xiongFundamentalAspectsQuantum2025, lorenzisEntanglementClassicalSimulability2025}. As in previous works \cite{liu2025, gong2025, Nerenberg:25, Sakurai:25, cimini2025}, the raw data $\mathbf{x}^{\text{raw}}_j$ is first classically preprocessed. In our implementation, input preprocessing consists of normalizing data to the interval $[0,1]$, while for image-classification tasks we additionally apply Principal Component Analysis (PCA) for dimensionality reduction of each raw data $\mathbf{x}^{\text{raw}}_j\in \mathbb{R}^{d_r}$ to $\mathbf{x}_j\in \mathbb{R}^{d_x}$, with $d_x \le d_r$. The resulting feature vector $\mathbf{x}_j$ is then encoded into the optical circuit via both the squeezing operations and the interferometer, so the final quantum state before measurement reads
\begin{equation}
   \ket{\psi_{out}(\mathbf{x}_j)} =  \hat{U}(\mathbf{x}_j)\hat{S}(\boldsymbol{r}(\mathbf{x}_j), \boldsymbol{\phi}(\mathbf{x}_j))\ket{\mathbf{0}}
\end{equation}
with input vacuum state $\ket{\mathbf{0}} = \bigotimes_{_i=1}^M\ket{0}$ and where 
\begin{equation}
    \hat{S}(\boldsymbol{r}(\mathbf{x}_j), \boldsymbol{\phi}(\mathbf{x}_j)) = \bigotimes_{k=1}^M \hat{S}_k(r_k(\mathbf{x}_j), \phi_k(\mathbf{x}_j))
\end{equation}
is the full squeezing transformation and $\hat{U}(\mathbf{x}_j)$ corresponds to the input-dependent interferometer matrix. 

On the one hand, classical data $\mathbf{x}_j$ to be processed are encoded in both squeezing amplitudes and angles \cite{nokkala2021} of the input modes via $\hat{S}(\boldsymbol{r}(\mathbf{x}_j), \boldsymbol{\phi}(\mathbf{x}_j))$ according to
\begin{equation}
\begin{aligned}
    \mathbf{r}(\mathbf{x}_j) &= \mathbf{r}_0 + \Delta\mathbf{r}(\mathbf{x}_j) \\
    \boldsymbol{\phi}(\mathbf{x}_j) &= [\boldsymbol{\phi}_0 + \Delta\boldsymbol{\phi}(\mathbf{x}_j)] \;\text{mod}\;2\pi
\end{aligned}
\end{equation}
where $\mathbf{r}_0, \mathbf{\phi}_0 \in\mathbb{R}^M $ are taken from uniform distributions $\mathcal{U}(a, b)$ of random numbers on $[a,b]$, $\mathbf{r}_0 \sim \mathcal{U}(0.5, 1.0)$ and $\mathbf{\phi}_0\sim\mathcal{U}(0, 2\pi)$. The input-dependent deviation for $\mathbf{a}\in\lbrace\mathbf{r},\boldsymbol{\phi}\rbrace$ is obtained as
\begin{equation}
    \Delta\mathbf{a}(\mathbf{x}_j) = 2\epsilon(\mathbf{x}^a_j - 0.5)\in[-\delta, \delta].
    \label{ec:delta encoding}
\end{equation}
The parameter $\epsilon$ is set to a fixed value, $\epsilon= 0.1$, throughout this work. The notation $\mathbf{x}^a_j$ indicates that different input components may be assigned to the squeezing strengths and phases, depending on the number of classical features per sample. This encoding allows up to $2M$ classical components to be introduced by assigning distinct values to the strength and phase of each optical mode.

On the other hand, input information is also introduced in the interferometer matrix $\hat{U}(\mathbf{x}_j)$ to further enhance the model's non-linearity. We employ the Cosine–Sine (CS) decomposition \cite{TangTian2024, suttonComputingCompleteCS2009} to enforce unitarity of the interferometer matrix $U$. This expresses any unitary matrix in terms of unitary factors and diagonal cosine ($C$) and sine ($S$) matrices satisfying $C^2 + S^2 = \mathbb{I}$. The classical input $\mathbf{x}_j$ is partially or fully embedded in the diagonal elements of $C$ and $S$, depending on the number of input features. In particular, this encoding strategy allows up to $\lfloor \frac{M}{2}  \rfloor$ classical components to be encoded, where $\lfloor \cdot \rfloor$ is the integer part. Further details are provided in App.~\ref{ap: Cosine-Sine Decomposition}. Note that this encoding in the interferometer allows for encoding at most half of the optical modes.

Combining both input and interferometer embeddings, the number of  encoded classical features $n_{\mathrm{cl}}$ scales linearly with the number of optical modes, according to $n_{\mathrm{cl}} = 2M + \left\lfloor \frac{M}{2} \right\rfloor$.  In particular, for an interferometer with $M=5$ modes, this corresponds to $n_{\mathrm{cl}} = 12$ classical parameters. The specific assignment of classical features to the circuits' parameters depends on the nature of the data and is detailed in Appendix~\ref{ap:input_encoding_details} for each task we consider later.

\subsection{Measured Observables}
Beyond the input encoding and the interferometer,  measurements at the output are also crucial to defining the quantum feature map. The choice of observables $\{\hat O_i\}$ is indeed a key design element, as it determines the structure and expressivity of the induced nonlinear mapping from classical data to output features. 

A given measurement scheme defines a quantum feature map whose components are given by expectation values of the selected observables,
\begin{equation}
O_{ij}(\mathbf{x}_j)
=
\bra{\psi_{\mathrm{out}}(\mathbf{x}_j)}\hat O_i\ket{\psi_{\mathrm{out}}(\mathbf{x}_j)}.
\label{eq:general_observation_matrix}
\end{equation}
The collection of these quantities for all samples gives the observation matrix. The different feature choices we have explored are described below.

\paragraph{Fock-basis sampling probabilities.}

Since the quantum device employs a GBS architecture, photon counting in the Fock basis is a natural measurement choice. We therefore consider projectors onto Fock states as observables. Let $ \mathcal {F} _ {N_p} $ denote the set of Fock configurations with total photon number $N_p$ as
$\mathcal{F}_{N_p} = \{\mathbf n \in\mathbb{N}^M: \sum_m n_m = N_p\}$. 
We define the projector-valued $\hat\Pi _{\mathbf{n}} = \ket{ \mathbf{n}} \bra{\mathbf{n}}$ observable family
\begin{equation}
\hat{\boldsymbol O}^{\Pi_\mathcal{P}}
=
\Big(
\hat \Pi _{\mathbf{n}}
\Big)_{\mathbf n \in \mathcal{F}(\mathcal P)},
\;
\mathcal{F}(\mathcal P)
=
\bigcup_{N_p \in \mathcal P}
\mathcal{F}_{N_p},
\label{eq:O_prob_multiNp_concat}
\end{equation}
with $\mathcal{P}$ be a finite set of total photon numbers. The resulting feature vector is given by the corresponding expectation values by concatenating the sampling probabilities over all photon-number sectors in $\mathcal P$. For example, the case $\mathcal{P} = \lbrace 2,4\rbrace$ corresponds to using all output multiple photon detection patterns comprising $2$ and $4$ photons in total. These probabilities admit analytic expressions in terms of Hafnians of matrices determined by the interferometer and squeezing parameters \cite{detailedGBS}. Including multiple photon-number sectors probes different Hafnian orders, enriching the nonlinear structure of the resulting feature map. Note that this measurement scheme would require PNR detectors. 

\paragraph{Photon-number moment observables.}
For benchmarking purposes, we also consider feature maps based on moments of the photon-number operators $\hat n_i$. Including first and second diagonal moments corresponds to the observable family 
\begin{equation}
\hat{\boldsymbol O}^{ n_i,n_i^2}
=
\big(\hat n_1,\dots,\hat n_M,\;\hat n_1^{\,2},\dots,\hat n_M^{\,2}\big).
\label{eq:O_mean_var_list}
\end{equation}
In practice, the second moments are used to compute the photon-number variances
$
\mathrm{Var}(\hat n_i)=\langle \hat n_i^{\,2}\rangle-\langle \hat n_i\rangle^2
$,
which are employed together with the mean photon numbers as features.

To further enhance expressivity, we also include inter-mode correlations by considering the full second-order moment family
\begin{equation}
\hat{\boldsymbol O}^{ n_i, n_in_j}
=
\hat{\boldsymbol O}^{ n_i,n_i^2}
\cup
(\hat n_a\hat n_b)_{1\le a<b\le M}.
\label{eq:O_second_order_list}
\end{equation}
This observable set captures pairwise photon-number correlations between optical modes, providing access to nonlocal information beyond single-mode statistics.

\paragraph{Quadrature covariance observables.}
Finally, beyond the GBS architecture, we also consider Gaussian measurements and the covariance matrix of the quadrature operators,
\begin{equation}
\hat{\boldsymbol O}^{\mathrm{q_ip_j}}
=
\left(
 \frac{1}{2}\{\hat R_i,\hat R_j\}
\right)_{i\le j},
\qquad
\hat{\boldsymbol R}=(\hat Q_1,\hat P_1,\dots),
\end{equation}
with $\lbrace \cdot, \cdot\rbrace$  anticommutator,  
which fully characterize the Gaussian state through the second-order moments, as the quadrature averages (first-order moments) vanish. 

Moment-based feature maps can be interpreted as coarse-grained descriptions of the underlying Fock-basis distribution, since photon-number operators and their products can be expressed as linear combinations of Fock-state projectors. Consequently, sampling-probability features provide a richer representation than moment-based ones, motivating their use as feature representation in our QELM. All these different sets of observables lead to different numbers of quantum features (see App.~\ref{ap: feature scaling}), thereby displaying different expressivity. 

\begin{figure*}[t]
    \centering
    \includegraphics[width=\textwidth]{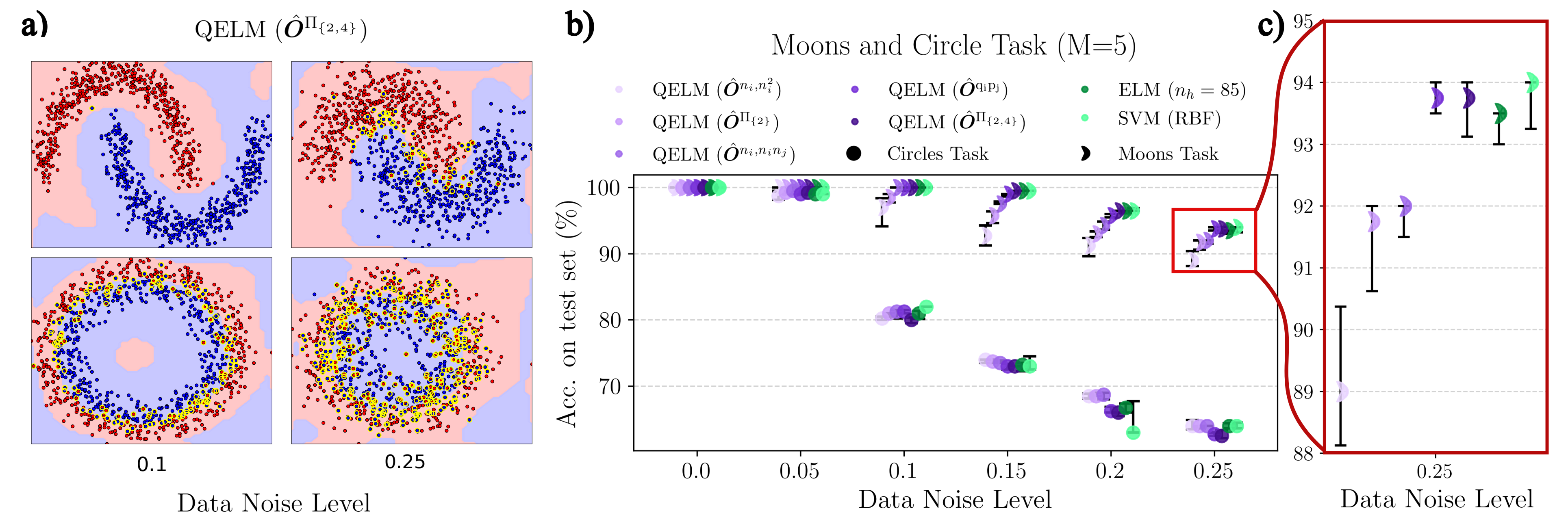}
    \caption{
    \textbf{a) Decision Boundaries:} QELM using Fock probabilities on the two-moon (top) and concentric circles (bottom) binary classification tasks. The results are shown for data noise levels (0.1 and 0.25), showcasing the model's ability to maintain separation despite increasing noise. The background colour corresponds to the model prediction, while the coloured dots indicate the correct class. Dots with a yellow border correspond to misclassified points. 
    \textbf{b) Test Set Accuracy vs. Noise Level:} This panel plots the classification accuracy on the test set against the noise level for all classical and quantum learning models on the Circles and Moons tasks. Points are the median across $10$ repetitions with different random seeds, and error bars correspond to the 25th and 75th percentiles. From left to right, the quantum models correspond to increasing output feature dimensionality (see App.~\ref{ap: feature scaling}).
    \textbf{c) Zoomed Accuracy for High Noise:} This inset magnifies the results for a noise level of 0.25 for the moons task.}
\label{fig:moons_and_circles}
\end{figure*}

\subsection{Classical Output Layer and benchmarks}
Once the quantum features are extracted from the GBS device, we apply feature-wise normalization, a standard procedure to stabilize the subsequent training process. The normalized features are then passed to a classical Ridge regression classifier   
\cite{scikit-learn}, as illustrated schematically in Fig.~\ref{fig:scheme}. Because Ridge regression has a closed-form solution, the output weights can be computed directly from the measured features, avoiding iterative optimization (see App.~\ref{ap: Readout Training}). This reflects a key advantage of both QELM and classical ELM frameworks: by limiting training to a simple linear readout, the overall optimization effort is drastically reduced, thereby lowering the required computational resources. Note that for classification, we classify each data through the index of the largest component of the final prediction $\mathbf{y}_j^{\mathrm{pred}}$ obtained from the Ridge regression.  

A key challenge in quantum machine learning is the fair benchmarking of quantum models against classical baselines. This motivates comparisons beyond simple reference models, such as random guessing, linear classifiers, and related approaches commonly adopted in the literature \cite{Nerenberg:25, cimini2025}. In this work, we compare the QELM with a classical ELM consisting of a random hidden layer with $n_h$ units followed by Ridge regression, providing a natural classical analogue. We also include a nonlinear support vector machine (SVM) with a radial basis function (RBF)  kernel as a strong reference \cite{scikit-learn}. Unlike the ELM and QELM, whose number of trainable parameters is fixed by the output feature dimension, SVM training is computationally more expensive, relying on iterative optimization with a variable number of trainable parameters. 
For these fully classical models, the input features correspond to the preprocessed data $\mathbf{x}_j$, so all quantum and classical models have the same input.

\section{Results}
The performance of the proposed QELM is assessed on both synthetic classification benchmarks and image-based tasks. The synthetic datasets are processed by simulating a GBS device with $M=5$ optical modes, while image classification is performed with $M=12$ modes.
The number of classical features that can be encoded is constrained by the available optical modes, yielding $n_{\mathrm{cl}}=12$ for $M=5$ and $n_{\mathrm{cl}}=30$ for $M=12$ within the adopted encoding scheme (see App.~\ref{ap:input_encoding_details}). For each repetition, a simple $5$-fold cross-validation is used to set the classical readout hyperparameters, such as the Ridge regularization strength and the RBF-SVM hyperparameters. Details of the simulation methods are provided in App.~\ref{ap:simulation}. The code for reproducing the results is available in a GitHub repository, which will be made public upon publication.

\subsection{Synthetic Tasks}
As an initial benchmark to assess the capabilities of the proposed QELM model, we consider two standard synthetic binary classification problems: the two moons and concentric circles datasets. In both cases, the task is to classify two-dimensional data points based on their positions in the plane. For both tasks, 800 data points are used for training, and 200 for testing. 

We also assess the model under noisy input conditions by adding Gaussian noise to the input data. This noise perturbs the position of each point, increasing the overlap between the two classes and thus the difficulty of the classification task. The results are presented in Fig.~\ref{fig:moons_and_circles}, where the full $100\%$ performance achieved by all quantum and classical architectures (QELMs, ELM, and SVM) is observed only for low levels of input noise. The data noise level corresponds to the standard deviation of the Gaussian perturbation applied during data generation. For the classical ELM, the number of hidden neurons is chosen to match the number of quantum features produced by the QELM using Fock probabilities with $\mathcal{P}=\lbrace 2,4 \rbrace$, which yields the largest feature dimension (see App.~\ref{ap: feature scaling}). 

As illustrated in Fig.~\ref{fig:moons_and_circles}\textbf{a)}, the QELM preserves a meaningful separation between classes even in the high-noise regime, indicating that the underlying quantum feature map retains sufficient discriminative capacity in this noisy setting. However, at high noise levels, it is impossible in principle to achieve perfect accuracy, as points from different classes begin to overlap significantly.  This is clearly seen in Fig.~\ref{fig:moons_and_circles}\textbf{b)}, where all models exhibit a similar decrease in performance as noise grows. 

Overall, these results show that good performance can be achieved with only $M=5$ optical modes and a minimal training procedure, with QELM accurate as the more demanding SVM. While these benchmarks do not reveal a clear separation between classical and quantum approaches, they demonstrate that the proposed QELM is both expressive and robust to noisy data, exhibiting behaviour that is approximately equivalent to that of its classical counterparts. 

Comparing performance across the different quantum models for the moons task, we observe that models with lower output feature dimension (e.g for QELM $O^{\Pi_{\lbrace 2,4 \rbrace}}$ the output layer consist of $85$ features when $M=5$) exhibit a more abrupt degradation in accuracy as noise increases, as shown in Fig.~\ref{fig:moons_and_circles}\textbf{c)}. This behaviour can be directly understood in terms of training resources: a larger quantum feature space implies more trainable parameters in the output Ridge regression, which in turn enhances expressivity and improves generalization. This observation highlights the importance of explicitly accounting for the number of features and trainable parameters when comparing different models and interpreting performance differences. A similar training strategy is employed for the classical ELM; however, achieving a comparable parameter count requires substantially more internal neurons than optical modes (in this case $n_h=85)$. This contrast highlights the potential of quantum feature maps to generate high-dimensional representations from compact physical resources and underscores the relevance of parameter-efficient representations and training resources. In the following section, we systematically examine the role of feature dimensions and trainable parameters in more demanding settings such as image classification.

\subsection{Image Classification}

Having established the capabilities of the proposed QELM in a simple scenario, we move to a more complex task such as image classification. In particular, we will address the ten-digit MNIST task \cite{lecun2010mnist},
 a standard benchmark in machine learning, and the OrganC dataset from the medMNIST collection \cite{MedMNIST2023}, which presents a medical image classification problem. For both tasks, we use the full training and test data sets. We employ an interferometer with $M=12$ optical modes and perform PCA retaining $30$ classical components. The specific allocation of these components across the squeezing parameters and the interferometer unitary follows the encoding scheme detailed in App.~\ref{ap:input_encoding_details}. 

\begin{figure}[t]
    \centering
    \includegraphics[width=\columnwidth]{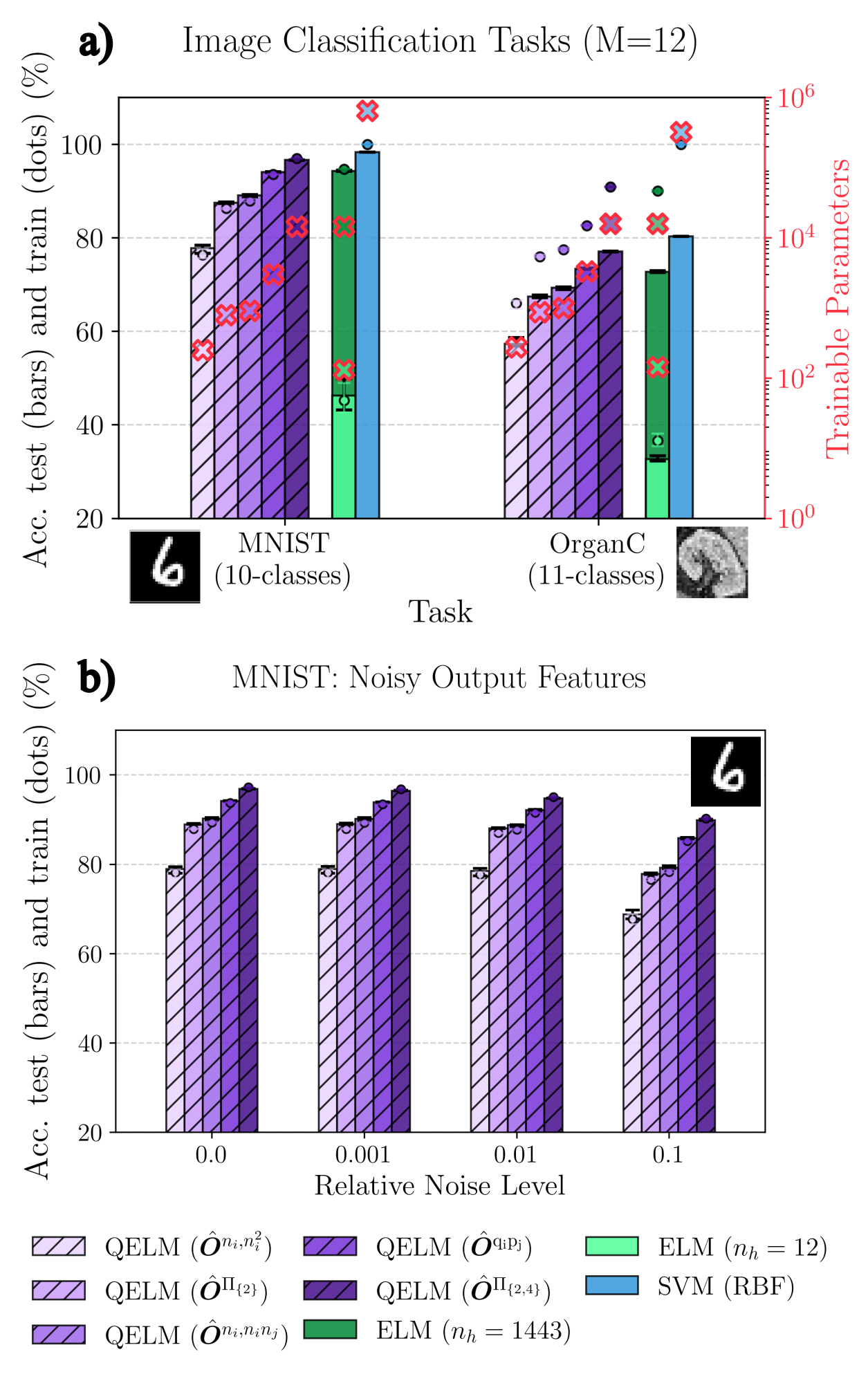}
    \caption{\textbf{a) Comparative Performance and Resource Analysis for Different Quantum and Classical Models (see color legend):} Presents the test(bars)/train(dots) set accuracy (left $y$-axis) and number of trainable parameters (right $y$-axis,  red cross markers) for two different image classification datasets (MNIST, OrganC). Points/bar height are the median across $10$ repetitions with different random seeds, and error bars correspond to the 25th and 75th percentiles. From left to right, the quantum models correspond to increasing output feature dimensionality (see App.~\ref{ap: feature scaling}).
    \textbf{b) Noisy Quantum Features Performance:} Accuracies of quantum models when Gaussian random noise is introduced to the quantum features after measurement.}
    \label{fig:image_class}
\end{figure}

Performance across all QELMs and classical models is shown in Fig.~\ref{fig:image_class}\textbf{a)} for both tasks, where a trend similar to that observed in the binary classification setting emerges. We also include the number of trainable parameters used in each setting (red cross symbols). Quantum models with a higher feature dimension achieve higher accuracies, reaching performance comparable to or exceeding that of the classical models considered. In particular, we observe that the QELM $O^{\Pi_{\lbrace 2,4\rbrace}}$, with the same number of output features, outperforms the classical ELM in both tasks. This demonstrates the ability of the proposed QELM architecture to move beyond simple synthetic datasets and handle more realistic scenarios. For the OrganC task, we notice that all models exhibit mild overfitting (see difference between train and test performance, as the dot and bar are not at the same height); indeed, the training accuracy exceeds the test accuracy, as also reported in previous works \cite{Nerenberg:25}.

In the MNIST task, we also assess the model’s performance accounting for sampling noise, as shown in Fig.~\ref{fig:image_class}\textbf{b)}. Specifically, we simulate output noisy conditions by adding independent, zero-mean Gaussian perturbations to each feature, with a feature-dependent standard deviation set by the feature’s robust spread (the difference between its 95th and 5th percentiles) scaled by the chosen noise strength. The results demonstrate that the model is robust to such noisy readouts, maintaining high accuracy even at a relative noise level of $0.1$, corresponding to fluctuations of about $10\%$ of each feature’s variability. 

Finally, we highlight the role of different encoding strategies by showing that the same information can yield different performance depending on how it is embedded in the optical circuit. To isolate this effect, we restrict the input to the first six PCA components, which is the maximum supported by uniquely employing the CS interferometer encoding for $M=12$ modes. We benchmark four strategies that encode the same features in different circuit degrees of freedom: squeezing amplitudes only ($E^{S_r}$), squeezing phases only ($E^{S_\phi}$), CS unitary parameters only ($E^{U}$), and a mixed encoding that combines both ($E^{U,S_r,S_\phi}$). The results, shown in Fig.~\ref{fig:image_encodings}, reveal a clear advantage of CS unitary encoding over input-dependent squeezing alone. This gain is most pronounced for quantum models with lower feature dimension, where limited readout capacity makes performance more sensitive to the quality of the feature map. For instance, the mean photon observable $\hat{\boldsymbol O}^{ n_i,n_i^2}$ shows an improvement of $\sim5\% $ and the quadrature covariance features $\hat{\boldsymbol O}^{\mathrm{q_ip_j}}$ improve $\sim 10\%$ when using the CS unitary encoding compared to only squeezing encoding.

\section{Resources and Expressivity}
\subsection{Classical Baselines}
A fair comparison between quantum and classical models is not straightforward, since the notion of ``resources'' differs across architectures. A sizeable resource is the number of trainable parameters in the readout. From the results in Fig.~\ref{fig:image_class}\textbf{a)} (which reports accuracy together with the corresponding parameter count on the right axis), the nonlinear SVM achieves slightly higher accuracy than the best QELM model, but this comes with substantially more trainable parameters and an iterative optimization procedure that relies on repeated kernel evaluations. By contrast, QELM training reduces to Ridge regression with a closed-form solution, which scales more favorably with the number of features.

While classical and quantum ELM require similar training resources,  a classical ELM needs en exponentially larger number of neurons to match the full quantum feature space of a QELM. When the ELM has enough internal neurons to match the dimensionality of the QELM feature space, both approaches achieve comparable performance while constraining the ELM to use only $12$ neurons results in a clear drop in accuracy.  Thus, the potential practical advantage of the QELM lies not in an inherent quantum expressivity advantage (since classical ELMs with matched feature dimension perform similarly), but rather in the ability to produce rich feature vectors from a compact, lower size physical system.

\begin{figure}[t]
    \centering
    \includegraphics[width=\columnwidth]{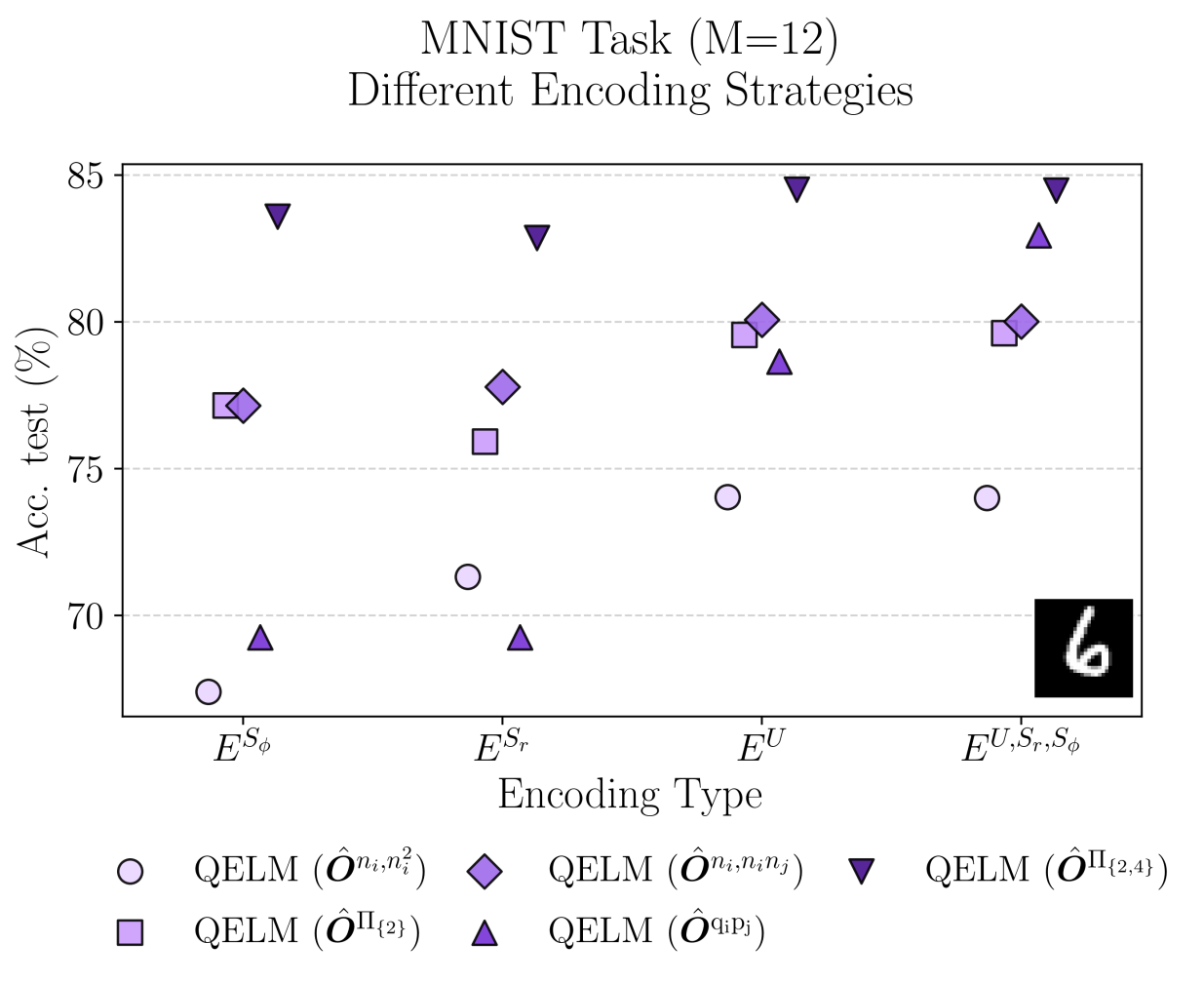}
    \caption{\textbf{Different encoding strategies:} Performance Comparison on MNIST task using only the first six PCA components to encode information in the GBS device with different strategies. $E^{S_\phi},E^{S_r},E^{U}$ corresponding to encodings employing only the squeezing phases, or  squeezing strengths, or the interferometer CS encoding, respectively. Values correspond to the median across repetitions. Error bars are not shown for clarity, but they are small in all cases.}
    \label{fig:image_encodings}
\end{figure}

\subsection{Measurement Scheme Effects}
The choice of measured features has a key role in QELM. Comparing different output choices, we find that the dominant factor behind observed performance differences is the dimensionality of the extracted quantum feature space, which directly determines the number of trainable parameters in the Ridge readout. In the moons task (Fig.~\ref{fig:moons_and_circles}\textbf{c)}), models with lower feature dimension exhibit a more abrupt degradation in accuracy as noise increases. Enlarging the feature space yields a richer representation of the data, which generally enhances both expressivity and robustness to noise, provided that overfitting is properly controlled. This trend is reinforced by the classification results in Fig.~\ref{fig:image_class}\textbf{a)}, where accuracy increases systematically with the feature dimension obtained from the measurement design (proportionally related to red cross symbols). Importantly, within the QELM family, this increase in trainable capacity is achieved without modifying the underlying physical resources, since all optical implementations use the same interferometer size. Instead, the feature dimension is controlled by the choice of measured observables.  

In particular, higher-order Fock space observables enlarge the feature space and improve performance. Among all the sets of observables considered, the best quantum model is obtained from Fock sampling with  $\mathcal{P}=\lbrace 2,4 \rbrace$, as observed in Figs.~\ref{fig:moons_and_circles} and \ref{fig:image_class}. This measurement scheme yields the largest set of quantum features $D^{\Pi_{2,4}}=1443$ ($M=12$) as detailed in App. \ref{ap: feature scaling}. At the same time, the second-best quantum model is based on quadrature covariance matrix features, which fall within the class of Gaussian processes that can be simulated efficiently on classical computers \cite{mari2012} with output feature dimension $D^{\mathrm{q_i p_j}}=300$ ($M=12$). Therefore, for the tasks considered here, it is difficult to attribute the observed gains uniquely to access to GBS samples, since comparable performance is also achieved using features that do not rely on non-Gaussian output statistics. However, the performance of the quadrature-based model is strongly dependent on the quantum design as it relies heavily on the interferometer encoding strategy. As shown in Fig.~\ref{fig:image_encodings}, the same quadrature features yield substantially lower accuracy when the input is encoded only through squeezing parameters, underscoring that the encoding itself plays a pivotal role in extracting expressive feature maps.

\subsection{Comparison with Reported Performances}
    
\begin{table*}[t] 
    \centering
    \begin{tabular}{lccccc}
        \toprule
        Model & Modes & Measured Output & Light Source & Optimization & Test Accuracy (\%) \\
        \midrule

        $\hat{\boldsymbol O}^{\Pi_{\{2,4\}}}$ (ideal)
        & $12$ & 2-,4-photon PNR probs. & Squeezed & Analytic (Lin.\ Reg.) & $96.92(0.07)$ \\

        $\hat{\boldsymbol O}^{\Pi_{\{2,4\}}}$ ($1\%$ noise)
        & $12$ & 2-,4-photon PNR probs. & Squeezed & Analytic (Lin.\ Reg.) & $94.81(0.05)$ \\

        $\hat{\boldsymbol O}^{\Pi_{\{2,4\}}}$ ($10\%$ noise)
        & $12$ & 2-,4-photon PNR probs. & Squeezed & Analytic (Lin.\ Reg.) & $89.93(0.16)$ \\

        \midrule
        \multicolumn{6}{l}{\textit{Related numerical results}} \\
        \midrule

        \textit{Nerenberg et al.} \cite{Nerenberg:25}
        & $10$ & $\le$ 4-photon PNR probs & Fock & Analytic (Lin.\ Reg.) & $\sim 88.0$ \\

        \textit{Sakurai et al.} \cite{Sakurai:25}
        & $12$ & 3-click probs. & Fock & Iterative (Lin.\ clf., AdaGrad) & $95.69$ \\

        \textit{Rambach et al.} \cite{rambach2025}
        & $20$ & 3-click probs. & Fock & Iterative (Lin.\ clf., AdaGrad) & $96.6$ \\

        \midrule
        \multicolumn{6}{l}{\textit{Experimental works}} \\
        \midrule

        \textit{Rambach et al.} \cite{rambach2025}
        & $12$ & 3-click probs. & Fock & Iterative (Lin.\ clf., AdaGrad) & $\sim 94.0$ \\

        \textit{Cimini et al.} \cite{cimini2025}
        & $>400$ & $\langle n_i\rangle,\ \mathrm{Cov}(n_i,n_j)$ & Squeezed & Iterative (Lin.\ SVM) & $\sim 60.0$ \\

        \textit{Gong et al.} \cite{gong2025}
        & $8176$  & $\sim 2200$ clicks$^\ast$ & Squeezed & Analytic (Lin.\ Reg.) & $95.86$ \\

        \bottomrule     
    \end{tabular}
    \caption{Comparison of photonic QELM models' resources for the MNIST task. The noisy values correspond to those in Fig~\ref{fig:image_class}. Accuracies from our models are reported as medians over 10 trials; values in parentheses give half the interquartile range in percentage points. $^\ast$ After selecting spatial–temporal modes, the measured output in \textit{Gong et al.} \cite{gong2025} selects the most frequent output events, instead of the probabilities of 2-3-4 photon detections of expected number moments.}
    \label{tab:model_comparison}
\end{table*}
Over the last few years, several photonic implementations of QELM models have been reported, spanning different optical resources, light degrees of freedom, encoding strategies, detection models, and classical readout procedures. As a consequence, direct performance comparisons are not straightforward. In this section, we summarize recent related results for the MNIST task, including both numerical and experimental studies, as well as Fock and squeezed light approaches, as listed in Tab.~\ref{tab:model_comparison}. 

To our knowledge, this work presents one of the first numerical studies of a GBS-based QELM for image classification. Regarding accuracy, the best numerical results in Tab.~\ref{tab:model_comparison} lie in a relatively narrow range ($\sim94\textrm{-}97\%$), despite relying on rather different physical resources and detection models. In particular, our best-performing simulated model uses full PNR probabilities from the $\{2,4\}$-photon sectors, similar to the measurement method used by \textit{Nerenberg et al.} \cite{Nerenberg:25}, whereas the approaches of \textit{Sakurai et al.} \cite{Sakurai:25} and \textit{Rambach et al.} \cite{rambach2025} use bucket detection and post-selected multi-click coincidence distributions. These are experimentally simpler to access; therefore, accuracy alone does not fully capture differences in physical complexity or the structure of the extracted quantum features. 

Furthermore, a proper comparison should also account for how the classical information is embedded into each photonic device. In \textit{Nerenberg et al.}, the first 20 principal components are encoded in variable beamsplitters of a  random reservoir; in \textit{Sakurai et al.}, PCA reduces the image to $M$ components, which are then mapped to phases and injected between a random pre-circuit and a random reservoir interferometer; and \textit{Rambach et al.} follow the same general PCA-to-phase strategy in an integrated boson-sampling architecture. In the GBS setting, \textit{Cimini et al.} \cite{cimini2025} use the first 100 principal-component values as inputs to a frequency-domain reservoir through pump spectral-phase shaping, whereas \textit{Gong et al.} \cite{gong2025} map PCA features onto their large temporal-spatial device so that each feature is associated with a spatial mode and its value selects a temporal mode. These examples illustrate that the encoding stage is highly platform-dependent and, as observed in the present work, different ways of injecting the same classical information into the same device can lead to noticeably different performance. 

The classical readout is also not identical across the literature and we have shown that this is a critical aspect. \textit{Nerenberg et al.} and \textit{Gong et al.} use analytic training of the final linear layer similar to us, whereas \textit{Sakurai et al.} and \textit{Rambach et al.} employ closely related iterative multiclass linear classifiers trained with AdaGrad; \textit{Rambach et al.} refer to their model as an L-SVC, but operationally it plays the same role as the linear readout in \textit{Sakurai et al.} \textit{Cimini et al.}, in turn, report several linear-readout options and use a linear-kernel SVM for the classification tasks highlighted in their paper.

Finally, among the experimental results collected in Tab.~\ref{tab:model_comparison}, the trade-off between accuracy and physical resources is particularly relevant. \textit{Rambach et al.} report about $94\%$ accuracy on a 12-mode photonic processor with three-photon Fock inputs, whereas \textit{Gong et al.} reach $95.86\%$ using an 8176-mode temporal--spatial GBS device operating at approximately 2200 average photon clicks. Thus, while the latter achieves the highest experimental accuracy, it does so with a substantially larger device. Likewise, \textit{Cimini et al.} demonstrate a large-scale GBS reservoir with more than 400 frequency modes and show that correlation-based features can be more informative than mean fields alone \cite{cimini2025}.

Our best simulated model with just 12 modes attains an accuracy close to that reported in the largest GBS image-classification experiment \cite{gong2025}. This should not be interpreted as a direct hardware-level comparison, since our results are obtained under controlled numerical conditions, whereas the experimental studies are affected by loss, finite sampling, device imperfections, and other implementation constraints. Rather, the encoding and measurement strategy proposed here can guide future experiments, as a good performance is achieved with a comparatively compact optical setting, indicating a possible route toward more resource-efficient photonic implementations. At the same time, the architecture's main ingredients are experimentally plausible. Programmable squeezed light sources with tunable squeezing levels and phases have been demonstrated, including time multiplexed implementations and reconfigurable multimode squeezed light generation \cite{dutt2016,tomoda2023,Kouadou2023}. Moreover, arbitrary linear-optical transformations can be synthesized from beam splitters and phase shifters through the Clements decomposition \cite{clements2016a}, which has already been implemented in photonic experiments \cite{zhu2025} for a generative quantum model supporting the experimental plausibility of the proposed framework.

\section{Conclusions}
In this work, we propose a QELM classification framework based on a GBS protocol, in which classical information is jointly encoded in the squeezing parameters and the interferometer matrix, thereby enhancing nonlinear processing of input data. The model was benchmarked across several classification tasks, including both binary and multiclass settings, and under different noise scenarios affecting the input data and the extracted quantum features. Across all cases, the proposed approach demonstrates stable and competitive performance. Our encoding enables good accuracy on the MNIST tasks with a reduced number of optical modes, highlighting its potential for more efficient optical implementations. Compared with classical baselines, namely an ELM and a nonlinear SVM, the QELM achieves accuracy comparable to the best-performing classical model while requiring significantly fewer trainable parameters and reduced optimization overhead. Notably, training in the QELM is limited to an analytic Ridge regression at the output layer, avoiding the iterative optimization procedures required by the SVM. 

Distinct sets of measured observables have been compared within the QELM framework, leading to different levels of expressivity and overall performance. We find a clear correlation between classification accuracy and the number of quantum features produced by each measurement scheme, which directly increases the number of trainable parameters in the final linear readout. This trend is natural, since a larger parameter space typically grants the output layer greater flexibility. In particular, Fock probability data achieves the best performance among quantum models. At the same time, a more fundamental question is whether one can design measurement schemes that boost performance without increasing the feature dimension, and thus without inflating the number of trainable parameters. Identifying such observables remains an open challenge and a promising path for future work. In this direction, our results highlight the importance of the encoding design in continuous-variable quantum machine learning. Embedding data directly into the interferometer via CS decomposition consistently improves performance over squeezing-only encodings, especially in low-feature regimes, indicating a promising route toward more expressive yet hardware-efficient photonic learning models.

Overall, these results demonstrate that direct quantum‑classical comparisons based solely on test accuracy are insufficient. A meaningful assessment must also account for the feature dimension, the number of trainable parameters, the optimization cost, and classical simulability. By explicitly taking into account these factors, we provide a balanced perspective on the potential advantages and limitations of quantum-enhanced learning approaches in the context of near-term  optical applications and show that a compact GBS‑based QELM with joint input encoding can achieve competitive classification performance using few optical modes and analytic training. 

\section{Acknowledgments}
We acknowledge the Spanish State Research Agency, through the COQUSY project PID2022-140506NB-C21 and -C22, funded by MICIU/AEI/10.13039/501100011033 and by ERDF, EU; through the María de Maeztu project CEX2021-001164-M, funded by MICIU/AEI/10.13039/501100011033;  MINECO funds the project through the QUANTUM SPAIN project, and EU through the RTRP - NextGenerationEU within the framework of the Digital Spain 2025 Agenda. CSIC Interdisciplinary Thematic Platform (PTI+) on Quantum Technologies in Spain (QTEP+) is also acknowledged. The project that gave rise to these results received the support of a fellowship from ”la Caixa” Foundation (ID 100010434). The fellowship code is LCF/BQ/DR24/12080033.

\bibliography{references}

\appendix
\section{Cosine-Sine Decomposition} \label{ap: Cosine-Sine Decomposition}
\subsection{CS-based Unitary Construction}
As mentioned in the main text, part of the classical input information is encoded directly in the interferometer matrix $U$. Therefore, propagating the information non-linearly into the output probability of the photons in each optical mode.

To encode $\mathbf{x}_j = (x_{j;0}, x_{j;1}, \dots, x_{j;n_{\mathrm{cl}}}) \in [0,1]^{n_{\mathrm{cl}}}$ with $n_{\mathrm{cl}}$ classical components into the unitary matrix of the interferometer, we employ the Cosine--Sine (CS) decomposition~\cite{TangTian2024, suttonComputingCompleteCS2009}. In general, the number of input features may exceed the number of components that can be effectively encoded in the interferometer using this approach. Therefore, we define
$\mathbf{x}^u_j = (x_{j;0}, x_{j;1}, \dots, x_{j;d_u})$, with $d_u \le n_{\mathrm{cl}}$, as the subset of features that are actually encoded in the interferometer.

The amount of classical data that can be encoded in this way is limited by the size of these matrices, which is related to the number of optical modes in the photonic circuit. In particular, $d_u = \left\lfloor \frac{M}{2} \right\rfloor,$ where $\left\lfloor \cdot \right\rfloor$ refers to the integer function. This decomposition involves the cosine $C$ and sine $S$ square $d_u \times d_u$ diagonal matrices, whose elements lie in the range $(0,1)$ and satisfy $C^2 + S^2 = I$, where $I$ represents the identity. 

In the following, we describe the specific implementations that have been used in this work corresponding to $M=12\;(d_u=6)$ and $M=5 \;(d_u=2)$. For the general case of the CS decomposition, refer to \cite{TangTian2024}.

\subsubsection{Balanced CS Decomposition}
For the interferometer with $M=12$ optical modes, we have an even number of components, so the decomposition is balanced in the sense that all matrices involved have the same shape. The interferometer matrix $U \in \mathbb{C}^{2d_u \times 2d_u}$ is decomposed as
\begin{equation}
U = 
\begin{pmatrix}
V_1 & 0 \\
0 & V_2
\end{pmatrix}
\begin{pmatrix}
C & S \\
S & -C
\end{pmatrix}
\begin{pmatrix}
W_1 & 0 \\
0 & W_2
\end{pmatrix}.
\label{ec:ap_Cs_balanced}
\end{equation}
where $V_1, V_2, W_1, W_2 \in \mathbb{C}^{d_u \times d_u}$ are randomly generated unitary matrices. The real-valued input vector 
$\mathbf{x}^u_j$ is introduced in the $C,S \in \mathbb{C}^{d_u \times d_u}$ matrices following 
\begin{equation}
C = \operatorname{diag}(\sqrt{\mathbf{x}^u_j}), \quad 
S = \sqrt{I - C^2}.
\label{ec: ap_CS_matrices}
\end{equation}
which ensures that $C^2 + S^2 = I $. Expanding the block multiplication  from Eq.~\eqref{ec:ap_Cs_balanced} yields:
\begin{align}
T_{11} &= V_1 C W_1, &
T_{12} &= V_1 S W_2, \\
T_{21} &= V_2 S W_1, &
T_{22} &= -V_2 C W_2,
\end{align}
so that
\begin{equation}
U =
\begin{pmatrix}
T_{11} & T_{12} \\
T_{21} & T_{22}
\end{pmatrix}.
\end{equation}
This structure guarantees that the interferometer matrix is unitary when $V_i$ and $W_i$ are also unitary.

\subsubsection{Unbalanced CS Decomposition}
In the case of $M=5$ optical modes, the decomposition is unbalanced since the interferometer matrix has odd dimension. This means that the matrices involved will have different sizes. The interferometer matrix decomposition reads
\begin{equation}
U = 
\begin{pmatrix}
V_1 & 0 \\
0 & V_2
\end{pmatrix}
\begin{pmatrix}
C & S & 0 \\
-S & C & 0 \\
0 & 0 & 1
\end{pmatrix}
\begin{pmatrix}
W_1^\dagger & 0 \\
0 & W_2^\dagger
\end{pmatrix},
\end{equation}
where $V_1, W_1 \in \mathbb{C}^{2\times2}$ and $ 
V_2, W_2 \in \mathbb{C}^{3\times3}$ are again randomly generated unitary matrices. The cosine and sine matrices are defined in the same way as in Eq.~\eqref{ec: ap_CS_matrices}, but limited to only $d_u= \left\lfloor \frac{5}{2} \right\rfloor = 2$ classical components, so they are $2 \times2$ matrices. This unbalanced CS structure preserves unitarity by extending the smaller $C,S$ blocks with zeros and identities to match the $5\times5$ dimension.

\section{Input Encoding}
\label{ap:input_encoding_details}
Depending on the task, the number and arrangement of the classical features encoded in the GBS device vary. In this section, we specify the encoding arrays $\mathbf{x}_j^a$ , with $a\in\{r,\phi,u\}$, as introduced in Eq.~\eqref{ec:delta encoding} and App.~\ref{ap: Cosine-Sine Decomposition}. 

For the moons and circles tasks, the input data consists of bidimensional samples $\mathbf{x}_j = (x_{j;0}, x_{j;1})$. In this case, we employ an optical circuit with $M=5$ modes, and the corresponding encoding arrays are given by
\begin{equation}
    \begin{aligned}
        \mathbf{x}^u_j &= (x_{j;0}, x_{j;1}),\\
        \mathbf{x}^r_j &= (x_{j;0}, x_{j;1}, 0.5, x_{j;0}, x_{j;1}),\\
        \mathbf{x}^\phi_j &= (x_{j;0}, x_{j;1}, 0.5, x_{j;0}, x_{j;1}).
    \end{aligned}
\end{equation}
Here, the same features are used to encode both the squeezing strengths and the phases in several modes, which enhances the effective nonlinearity of the resulting quantum feature map. The constant entries are fixed to $0.5$, which corresponds to zero modulation according to Eq.~\eqref{ec:delta encoding}.

For the image classification task, each sample is characterized by a large number of pixels. To reduce the input dimensionality, we perform PCA and retain the first 30 components. The resulting input vector is $\mathbf{x}_j = (x_{j;0}, x_{j;1}, \dots, x_{j;29})
$. Using an interferometer with $M=12$ modes, the encoding arrays are defined as
\begin{equation}
    \begin{aligned}
        \mathbf{x}^u_j &= (x_{j;0}, x_{j;1}, \dots, x_{j;5}),\\
        \mathbf{x}^r_j &= (x_{j;6}, x_{j;7}, \dots, x_{j;17}),\\
        \mathbf{x}^\phi_j &= (x_{j;18}, x_{j;19}, \dots, x_{j;29}).
    \end{aligned}
\end{equation}
In this task, different feature subsets are used to encode the interferometer, squeezing strengths, and phases, respectively, allowing the quantum model to exploit a richer, more structured embedding of the classical data.

For the results shown in Fig.~\ref{fig:image_encodings}, only the first $6$ PCA components are retained, so that the input vector is
$\mathbf{x}_j = (x_{j;0}, x_{j;1}, \dots, x_{j;5})$.
For the mixed strategy $E^{U,S_r,S_\phi}$, the encoding arrays are defined as
\begin{equation}
    \begin{aligned}
        \mathbf{x}^u_j &= (x_{j;0}, x_{j;1}, \dots, x_{j;5}),\\
        \mathbf{x}^r_j &= (x_{j;0}, x_{j;1}, \dots, x_{j;5}, 0.5, \dots, 0.5),\\
        \mathbf{x}^\phi_j &= (x_{j;0}, x_{j;1}, \dots, x_{j;5}, 0.5, \dots, 0.5).
    \end{aligned}
    \label{eq:K6_encodings_arrays}
\end{equation}
In this case, the same six PCA features are used to encode the interferometer, while the squeezing strengths and phases are modulated with the same feature vector supplemented by constant entries. For the strategies $E^{S_r}$ and $E^{S_\phi}$, the interferometer is kept random, and only the squeezing strengths or phases are modulated, respectively, using the encoding vectors defined in Eq.~\eqref{eq:K6_encodings_arrays}. For the strategy $E^{U}$, no squeezing modulation is applied, and only the interferometer is encoded according to Eq.~\eqref{eq:K6_encodings_arrays}.

\section{Feature Scaling} \label{ap: feature scaling}
Depending on the measurement scheme introduced after the input encoding stage (see Sec.~\ref{sec:Input Encoding} in the main text), the dimensionality of the quantum feature space can vary significantly while keeping the same physical resources, namely the number of optical modes $M$. This directly impacts both the expressivity of the induced feature map and the number of trainable parameters in the output layer.

For the measurement observable choices employed in this work, the corresponding number of quantum output features $D^{\mathrm{Mea}}$ is given by
\begin{equation}
    \begin{aligned}
        D^{\mathrm{n_i,n_i^2}} &= 2M ,\\ 
        D^{\Pi_{\lbrace 2\rbrace}} &= \binom{M+2-1}{2} ,\\
        D^{\mathrm{n_i, n_i n_j}} &= 2M + \frac{M(M-1)}{2} ,\\
        D^{\mathrm{\mathrm{q_i p_j}}} &= M(2M+1) ,\\
        D^{\Pi_{\lbrace 2,4\rbrace}} &= \binom{M+2-1}{2} + \binom{M+4-1}{4} .
    \end{aligned}
    \label{ec:ap_features_scaling}
\end{equation}
where, for $\Pi_{\mathcal P}$ features,
$D^{\Pi_{\mathcal P}}=\sum_{N_p\in\mathcal P}\binom{M+N_p-1}{N_p}$,
while the remaining dimensions follow from counting independent correlation-matrix entries.

\begin{figure}[]
    \centering
    \includegraphics[width=\columnwidth]{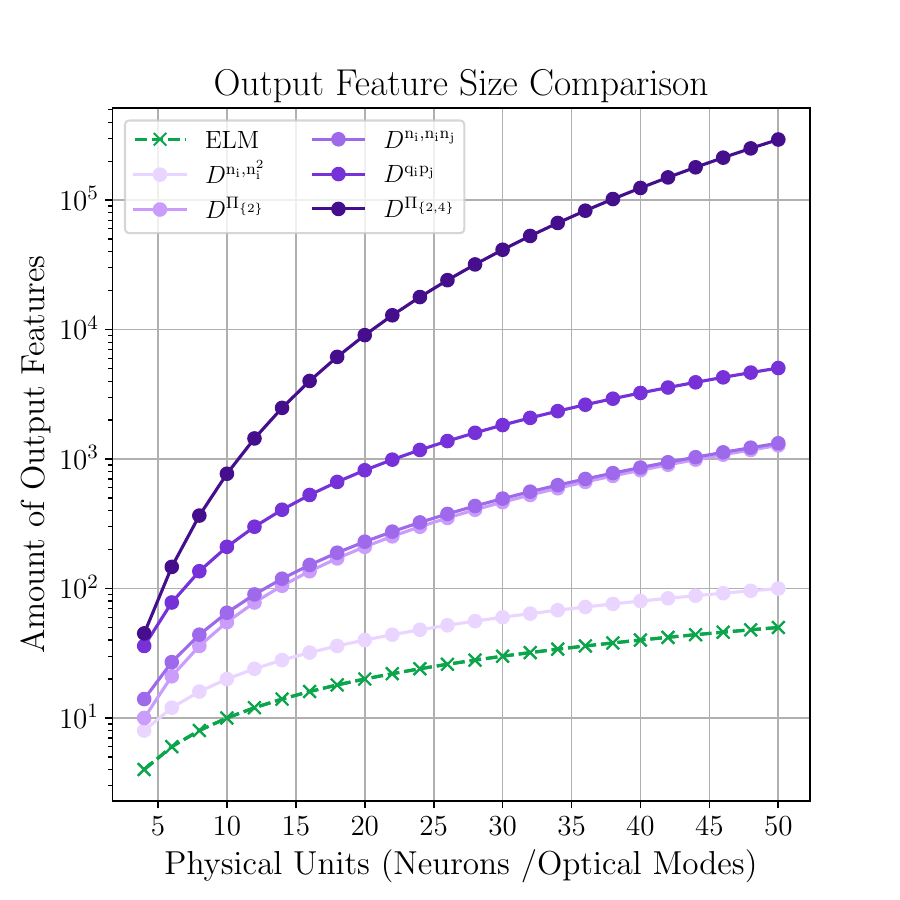}
    \caption{Scaling of the output feature dimension for the different measurement schemes considered in this work as a function of the number of optical modes $M$. For comparison, the linear scaling of the classical ELM feature space with respect to its number of internal neurons is also shown.}\label{fig:ap_feature_scaling}
\end{figure}

The resulting scaling of the feature dimension with the number of optical modes, as dictated by Eqs.~\ref{ec:ap_features_scaling}, is illustrated in Fig.~\ref{fig:ap_feature_scaling}. While moment-based measurement schemes lead to polynomial growth in $M$, Fock probability measurements exhibit a much steeper, combinatorial increase in the number of accessible features. As a reference, the linear scaling of the classical ELM feature space with its number of internal neurons is also reported.

This comparison highlights that, for a fixed number of optical modes, different quantum measurement strategies can generate feature spaces with vastly different dimensionalities. In particular, Fock probability measurement provides the largest feature spaces, which, in the benchmarks considered here, is associated with improved model performance, despite the underlying physical resources remaining unchanged. The concrete values for the interferometer sizes used in this work are shown in the table.~\ref{tab:feature_dims}.

\begin{table}[]
\centering
\vspace{0.5cm}
\setlength{\tabcolsep}{12pt}
\begin{tabular}{c|cc}
\toprule
 Optical Modes & $M=5$ & $M=12$ \\
\midrule
$D^{\mathrm{n_i,n_i^2}}$      & $10$  & $24$   \\
$D^{\Pi_{\lbrace 2\rbrace}}$  & $15$  & $78$   \\
$D^{\mathrm{n_i, n_i n_j}}$   & $20$  & $90$   \\
$D^{\mathrm{q_i p_j}}$        & $55$  & $300$  \\
$D^{\Pi_{\lbrace 2,4\rbrace}}$ & $85$  & $1443$ \\
\bottomrule
\end{tabular}
\caption{Feature-space dimension for selected values of $M$.}
\label{tab:feature_dims}
\end{table}
\section{Implementation and Numerical Details}
\subsection{Readout Training and Classical Baselines}\label{ap: Readout Training}

Regarding the QELM models, the measured observables, as defined in
Eq.~\eqref{eq:general_observation_matrix}, are concatenated to build the
feature matrix
$\mathcal{O}\in\mathbb{R}^{N^{\mathrm{train}}\times D^{\mathrm{obs}}}$,
where $N^{\mathrm{train}}$ is the number of training samples and
$D^{\mathrm{obs}}$ is the dimension of the output features, as in
App.~\ref{ap: feature scaling}. For a classification task with $C$ classes, the labels are encoded in a one-hot matrix $Y\in\mathbb{R}^{N^{\mathrm{train}}\times C}$. The classical readout is
trained by ridge regression, minimizing
\begin{equation}
\min_{W,\mathbf{b}}
\left\|Y-\mathcal{O}W-\mathbf{1}\mathbf{b}\right\|^{2}+\alpha\left\| W\right\|^{2},
\end{equation}
where $W\in\mathbb{R}^{D^{\mathrm{obs}}\times C}$ is the matrix of trainable weights, $\mathbf{b}\in\mathbb{R}^{1\times C}$ is the intercept vector, $\mathbf{1}\in\mathbb{R}^{N^{\mathrm{train}}\times 1}$ is a vector of ones,
and $\alpha$ is the regularization strength. In practice, this is implemented with \texttt{sklearn.linear\_model.Ridge} from \textit{scikit-learn} \cite{scikit-learn}. For a new input $\mathbf{x}_j$, the predicted output scores are
\begin{equation}
\mathbf{y}^{\mathrm{pred}}_j
=
\mathbf{o}(\mathbf{x}_j)W^{\mathrm{opt}}
+
\mathbf{b}^{\mathrm{opt}},
\end{equation}
where $\mathbf{o}(\mathbf{x}_j)\in\mathbb{R}^{1\times D^{\mathrm{obs}}}$ is the
output feature vector associated with $\mathbf{x}_j$. The predicted class is
then assigned as
\begin{equation}
\hat{c}(\mathbf{x}_j)
=
\arg\max_{c\in\{1,\ldots,C\}}
\left[
\mathbf{y}^{\mathrm{pred}}_j
\right]_c .
\end{equation}

For the classical baselines, the RBF-kernel SVM is implemented with the
\texttt{SVC} classifier from \textit{scikit-learn} \cite{scikit-learn}. The
one-hot targets are converted into class labels, and the model is fitted
directly on the input feature matrix using the \texttt{libsvm}-based
optimization implemented in \texttt{SVC}. The classical ELM uses a fixed random hidden layer, with weights and biases sampled from a normal distribution, followed by a sigmoid activation function. No optimization is performed in the hidden layer; only the final readout is trained, using the same ridge-regression output layer as in the QELM models.

\subsection{Simulation Details and Data Generation}\label{ap:simulation}

All numerical simulations of the GBS device are performed using the
\textit{Strawberry Fields} Python package \cite{Killoran_2019} with the
\textit{gaussian} backend. In this representation, the state is fully described by its phase-space mean vector and covariance matrix, from which mean photon numbers, photon-number correlations, and quadrature covariance matrices are computed directly. The Fock-basis probabilities used as features are evaluated from the same Gaussian-state description for the corresponding output occupation patterns, where all queried patterns satisfy $N_p\leq 4$. 

The data corresponding to the binary synthetic tasks are generated using \textit{scikit-learn} \cite{scikit-learn}, which also allows us to introduce the Gaussian noise used in the corresponding numerical experiments.

\end{document}